\documentclass[a4paper,11pt]{article}
\pdfoutput=1 %
\usepackage{jcappub} 

\usepackage[T1]{fontenc} 

\usepackage[utf8]{inputenc}

\usepackage{hyperref} 
\usepackage{url}
\usepackage{graphicx}
\usepackage{esvect}
\usepackage{graphics}
\usepackage{natbib}
\usepackage{mathrsfs}
\usepackage{blindtext}
\usepackage{comment} 
\usepackage{placeins} 

\usepackage[table]{xcolor}
\usepackage{multirow}

\usepackage[normalem]{ulem}
\usepackage{color}

\usepackage{soul}

\usepackage{supertabular}
\usepackage{tabularx}

\title{Cosmic-Ray Propagation Models Elucidate the Prospects for Antinuclei Detection}

\author[a, b]{Pedro~De~La~Torre~Luque}
\emailAdd{pedro.delatorreluque@fysik.su.se}
\author[c]{Martin Wolfgang Winkler}
\emailAdd{martin.wolfgang.winkler@gmail.com}
\author[b]{Tim Linden}
\emailAdd{linden@fysik.su.se}

\affiliation[a ]{Instituto de Física Teórica, IFT UAM-CSIC, Departamento de Física Teórica,\\
Universidad Autónoma de Madrid, ES-28049 Madrid, Spain}
\affiliation[b ]{The Oskar Klein Centre, Department of Physics, Stockholm University, AlbaNova\\
  SE-10691 Stockholm, Sweden}
\affiliation[c ]{Department of Physics, The University of Texas at Austin, Austin, 78712 TX, USA}
\date{\today}

\abstract{
Tentative observations of cosmic-ray antihelium by the AMS-02 collaboration have re-energized the quest to use antinuclei to search for physics beyond the standard model. However, our transition to a data-driven era requires more accurate models of the expected astrophysical antinuclei fluxes. We use a state-of-the-art cosmic-ray propagation model, fit to high-precision antiproton and cosmic-ray nuclei (B, Be, Li) data, to constrain the antinuclei flux from both astrophysical and dark matter annihilation models. We show that astrophysical sources are capable of producing $\mathcal{O}(1)$ antideuteron events and $\mathcal{O}(0.1)$ antihelium-3 events over 15~years of AMS-02 observations. Standard dark matter models could potentially produce higher levels of these antinuclei, but showing a different energy-dependence. Given the uncertainties in these models, dark matter annihilation is still the most promising candidate to explain preliminary AMS-02 results.
Meanwhile, any robust detection of antihelium-4 events would require more novel dark matter model building or a new astrophisical production mechanism.
}

\begin{document}
\maketitle

\section{Introduction}
\label{sec:intro}

Cosmic-ray antimatter has long been used to search for signatures of physics beyond the standard model~\cite{Silk1984, Stecker1985, 1988ApJ_Rudaz, Jungman, ELLIS1988403, Bottino1998, Bergstrom_1999, Donato2004, Fornengo:2013xda}, in particular for weakly interacting massive particles (WIMPs). However, careful observations have not uncovered convincing evidence of a WIMP signature in either positron or antiproton observations~\cite{Pamela_Antip, Pam_Antip2010, CIRELLI20091, Hooper_2015, John_2021, Ap_analysis}. In fact, astrophysical uncertainties have precluded our ability to conclusively observe or rule out thermal WIMPs as subdominant contributors to either channel~\cite{John_2021, Cholis_DM_e+_Bounds, Winkler:2017xor, Calore2021}. 

This ambiguity motivates searches for heavier cosmic-ray antinuclei, such as antideuterons or antihelium, which have significantly smaller astrophysical backgrounds. However, the predicted antinuclei flux from standard WIMP annihilation is also much smaller, making the detection of an antinuclei signal experimentally challenging. 

Remarkably, observations by the Alpha Magnetic Spectrometer (AMS-02) on board the international space station have tentatively detected approximately 10 antihelium events~\cite{Ting:2016}. The AMS-02 data also include several events with mass identification most consistent with antideuterons, though it is difficult to eliminate the possibility that these mass measurements constitute the tail of the antiproton distribution~\cite{Ting:2016, vonDoetinchem:2020vbj}. Even more surprisingly, AMS-02 reported that a handful of the observed antihelium events have a detected mass more consistent with antihelium-4, rather than antihelium-3 nuclei -- though it remains possible that all of the detected events are antihelium-3. The roughly similar (within an order of magnitude) flux of antihelium-4, antihelium-3 and antideuterium nuclei would be extremely unexpected. Kinematic considerations in secondary production make the addition of each additional antinucleon significantly less probable. Models predict that each successive antinucleus species should have a flux that is suppressed by $\sim\mathcal{O}(10^{4})$ if one antinucleon is added~\cite{vonDoetinchem:2020vbj}.

This has motivated the exploration of more exotic methods for producing heavy antinuclei. For antihelium-3, this has included models which enhance the probability that antinucleons ``coalesce'' into the more stable antihelium-3 nucleus~\cite{Carlson:2014ssa, Cirelli:2014qia}, as well as models where astrophysical processes re-accelerate antihelium-3 nuclei to higher energies where they are more detectable~\cite{Cholis:2020twh}. Most excitingly, we recently demonstrated that the off-vertex decay of the $\bar{\Lambda_b}$ particle is potentially capable of producing antihelium nuclei with high efficiency ~\cite{Winkler:2020ltd}. 
Whether these decays are able to explain the $\overline{He}$ signal depends on the branching ratio of the process $\overline{\Lambda}_b \rightarrow \overline{He}$, which is expected to be measured in accelerator experiments~\cite{ALICEWhite} in the near future.


However, none of these models is capable of producing a detectable antihelium-4 flux. In particular, kinematic considerations prevent $\overline{\Lambda}_b$ decays from producing any antihelium-4. This has motivated the investigation of models with extended dark sectors~\cite{AN_Cascade, Heeck:2019ego, Curtin2022}, or even novel cosmological models that include significant antimatter clouds~\cite{Poulin_Antiworld}, which currently comprise the only known methods to produce a bright antihelium-4 flux.

Regardless of the authenticity of these events, the fact that we are on the cusp of entering a data-rich age of cosmic-ray antinuclei observations motivates us to more carefully analyze the expected flux from both astrophysical and beyond standard model processes. In this paper, we utilize state of the art cosmic-ray production and propagation models, which are constrained by up-to-date observations 
of secondary nuclei (most importantly: B, Be, and Li), in order to predict both the astrophysical and dark matter induced fluxes of antideuterons and antihelium. We pay particular attention to uncertainties from nuclear cross-sections, as well as the effect of standard coalescence models on the expected antinuclei flux. Our work improves on previous studies by carefully implementing the present antiproton bounds on WIMP annihilation, which allows us to robustly constrain the antinuclei fluxes from dark matter.

Our models indicate that standard astrophysical mechanisms would be expected to produce $\mathcal{O}(1)$ antideuterium event that is detected by AMS-02 over a 15-year observation period, but only $\mathcal{O}(0.1)$ antihelium-3 event. Standard WIMP annihilation into bottom-quark pairs, on the other hand, is capable of producing $\mathcal{O}(1)$ antideuterium and $\mathcal{O}(1)$ antihelium-3 event, in optimistic cases, that are detectable by AMS-02 over 15~years. In order to explain the $\mathcal{O}(10)$ antihelium-3 events, indicated by the preliminary AMS-02 data, a somewhat larger branching ratio $\overline{\Lambda}_b \rightarrow \overline{He}$ compared to our baseline prediction with Pythia is required, or more general dark matter final states need to be considered. As expected, neither standard astrophysics nor our standard WIMP implementation produce any measurable fluxes of antihelium-4.
We finally discuss the most important systematics in the determination of events, and future astrophysical and collider experiments that would be capable of advancing the field.

This paper is organized as follows: We discuss our coalescence approach for antinuclei formation in Section~\ref{sec:Coalescence}. Then, we detail our computations for the spectra of light antinuclei (antideuterons and antihelium), using a new version of the {\tt DRAGON2} code, and explain the set-up employed to predict their fluxes at Earth in Section~\ref{sec:DRAGON}. In Sections~\ref{sec:astrop} and~\ref{sec:DM} we calculate the expected antideuteron and antihelium-3 spectra produced from CR interactions and WIMP annihilations, respectively, and compare our results with current instrumental sensitivities. In Section~\ref{sec:Maximal}, we predict the flux upper limit for both antinuclei, based on the antiproton DM bounds derived in our companion work. Finally, we summarise and discuss our main findings in Section~\ref{sec:conc}.

\section{Coalescence Modeling} 
\label{sec:Coalescence}
One of the most important ingredients in the calculation of antinuclei fluxes is the modeling of the coalescence process, i.e.\ the application of the condition under which one or several antinucleons merge.

Although experimental efforts are being made to improve our modeling of the coalescence~\cite{ALICE:2022zuz, ALICE2022}, it is extremely difficult to transfer our fundamental understanding of particle coalescence into an accurate prediction of the coalescence probability in any particle physics interaction. Thus, we instead employ simplified models that are fit to current data. In this formalism, the momentum and position of each antinucleon is compared to every other antinucleon produced in the interaction. If the antinucleons are ``close enough'' in this six-dimensional parameter space, then it is assumed that the antinucleus will efficiently form. 

\paragraph*{Event-by-Event Coalescence Model}
$\,$\\
A more detailed evaluation of the interactions and showers as modeled by an event generator, such as Pythia~\cite{Bierlich:2022pfr} or HERWIG~\cite{Bellm:2015jjp}, allow for an improved calculation of the process of antinuclei formation, but requires to check the coalescence condition on every antinucleon pair (or triplet, etc.).

Specifically, we need to check the relative momentum of each antiproton/antineutron pair in its center-of-mass frame within a generated event. An antideuteron is formed if the relative momentum is smaller than the coalescence momentum $p_c$. Antihelium-3 can be produced in two ways, directly through the coalescence of an antineutron and two antiprotons, or through the production of antitriton, which decays to antihelium-3 on astrophysical timescales. There are two similar procedures for computing antihelium-3 coalescence, one can either: (a) examine each $\bar{p}\bar{p}\bar{n}$ and $\bar{p}\bar{n}\bar{n}$ triplet and determining whether each pair in the arrangement has a coalescence momentum smaller than $p_c$, or (b) determine the center-of-mass frame of each potential antinucleus and determine if the relative momentum of each nucleon in the so-obtained center-of-mass frame is smaller than $2^{1/6} p_c/2$. The factor $2^{1/6}$ is needed such the coalescence volume in the analytical and event-by-event model match (see Ref.~\cite{Winkler:2020ltd}). 

In addition to the coalescence condition, we need to require antinucleons to originate from the same particle vertex (either the primary vertex or a common displaced vertex).

The event-by-event model is computationally challenging, especially for very rare interactions like heavy antinuclei formation. 

\paragraph*{Analytic Coalescence Model}
$\,$\\
A much simpler method -- the analytic coalescence model -- involves calculating the spectra of individual antiprotons and antineutrons, and assumes that the multi-antinucleon spectra are obtained from the product of single-nucleon spectra~\cite{CHARDONNET1997313, Coal}. This results in the following (Lorentz-invariant) expression for the antinucleus formation,
\begin{equation}
E_{\bar{A}}\frac{d^3N_{\bar{A}}}{dp_{\bar{A}}^3} = B_A R(x) \left(E_{\bar{n}}\frac{d^3N_{\bar{n}}}{dp_{\bar{n}}^3}\right)^Z \times \left(E_{\bar{p}}\frac{d^3N_{\bar{p}}}{dp_{\bar{p}}^3}\right)^N  ,
\label{eq:general}
\end{equation}
where $E_{i}\frac{d^3N_{i}}{dp_{i}^3}$ is the invariant differential yield of the particle $i$, A is the nucleus mass number, with proton number Z and neutron number N and the sub-index ${\bar{n}}$ and ${\bar{p}}$ stands for antineutrons and antiprotons, respectively. In order to reject unphysical antinuclei production (where the simulation merges a prompt and a displaced antinucleon) it is important to enter the prompt yields (i.e.\ those produced at the initial vertex) for $E_{i}\frac{d^3N_{i}}{dp_{i}^3}$. The term $R(x)$, with $x=\sqrt{s^2 + A^2m_p^2 + 2\sqrt{s}\tilde{E}_A}$ and with $\tilde{E}_A$ as the centre of mass product nucleus energy, is a phase space correction, needed to account suppression close to the kinematical threshold, defined in Ref.~\cite{Duperray:2003tv}.

Physically, this model simply states that an antinucleus with $A$ constituents will form whenever $A$ antinucleons are found within a small enough region of momentum space that is given by the coalescence parameter, $B_A$. This parameter is tunable for each antinucleus species, and it contains the crucial information about the probability of coalescence. It can also be rewritten as a function of the coalescence momentum $p_c$,
\begin{equation}
B_A \simeq \left(\frac{1}{8} \frac{4 \pi p_c^3}{3}\right)^{A-1} \frac{m_{A}}{m_{p}^Z m_{n}^N}
\label{eq:B2}
\end{equation}

Despite its simplicity, there is an immediate drawback of the analytic coalescence model: in contrast to the event-by-event model it neglects the correlations in the antinucleon spectra, i.e.\ it assumes that the momentum of each antinucleon in a particle collision is uncorrelated with the existence and momentum of every other antinucleon (except for the kinematical correlation which is taken into account through the suppression factor $R(x)$). In particular, it misses the hard correlation in the multi-antinucleon distribution which, for instance, results from the clustering of outgoing jets in a hadronic event. Furthermore, it entirely misses the production of antinuclei at the displaced decay vertex of a heavy hadronic resonance (most importantly a $\bar{\Lambda}_b$).

\noindent 

\paragraph*{Event-by-Event vs. Analytic Coalescence Model}
$\,$\\
In the case of dark matter annihilation, the analytic coalescence model turns out to perform very poorly. First, in dark matter annihilation, particles are typically produced in two back-to-back jets. Hence, hard correlations in the antinucleon production -- which are not taken into account in the analytic model -- turn out to be very important~\cite{Kadastik:2009ts}. In addition, the light antinuclei fluxes receive a large (potentially dominant) contribution from the displaced decays of $\bar{\Lambda}_b$ particles which is entirely missed in the analytic coalescence model~\cite{Winkler:2020ltd}. Therefore, the event-by-event coalescence model is the only viable option to calculate antinuclei spectra from dark matter annihilation. 

On the other hand, the astrophysical antinucleus background arises mainly from proton-proton and proton-nucleus collisions in the soft-QCD regime. In this regime it is notoriously difficult for Monte Carlo generators to predict antinucleon spectra accurately. In fact, it was shown (see e.g.~\cite{Kappl:2014hha,Kachelriess:2015wpa,Gomez-Coral:2018yuk}) that the measured antiproton spectra in low and intermediate energy proton-proton collisions deviate from the predictions of common Monte Carlo generators by up to an order of magnitude. In this light, the analytic coalescence model is the better option for predicting the astrophysical antinucleus fluxes. This is because in the analytic coalescence model (as defined by Eq.~\eqref{eq:general}) we can employ the analytic parameterization of antinucleon production spectra~\cite{Winkler:2017xor} which were fitted to a wide collection of accelerator data. At the same time, we verified that the hard correlations that are not included in the analytical model do (practically) not impact the astrophysical antinuclei fluxes below a few tens of GeV (i.e.\ in the energy range for which we need more accurate predictions; where astrophysical fluxes are important).\footnote{We emphasize that, while hard correlations in the antinucleon production play a minor role for predicting the astrophysical antinucleus fluxes, it is important to include the kinematic suppression close to threshold through the factor $R(x)$ in Eq.~\eqref{eq:general}.}




\paragraph*{Our Approach}
$\,$\\
Due to the arguments presented above, we evaluate the cross sections for antideuteron and antihelium-3 production from astrophysical processes (proton-proton, proton-nucleus and nucleus-nucleus collisions) in the analytic coalescence model (Eq.~\ref{eq:general}). For this purpose we employ the antiproton and antineutron cross section parameterization from Ref.~\cite{Winkler:2017xor}, which also accounts for isospin effects in the production of antineutrons. We only include the prompt contribution to the cross section, i.e.\ we do not include hyperon-induced antinucleons. The latter cannot merge into antinuclei because they are produced too far away from the primary vertex (outside the range of the nuclear force).

In contrast, for the DM case, we employ the event-by-event coalescence model. With the {\tt Pythia} 8.3 event generator~\cite{Bierlich:2022pfr} we simulate the annihilation of two incoming dark matter particles into bottom quarks through an intermediate colorless electrically-neutral resonance. Then, we apply the coalescence condition to any pair or triplet of antinucleons. 
For those antinucleons which successfully merge into an antinucleus we extract their momentum. In this way we can derive input tables with the antideuteron and antihelium spectra at production (dN/dE) for {\tt DRAGON2} (see Section~\ref{sec:DRAGON}).

In our computations, we account for the production of antinuclei from the displaced-vertex decay of $\overline{\Lambda}_b$ particles, as well as the correction of the transition ratio $f(b \rightarrow \overline{\Lambda}_b)$ pointed out by Ref.~\cite{Winkler:2020ltd} (see also Ref.~\cite{Comment}). This correction is necessary because $f(b \rightarrow \overline{\Lambda}_b)$ is systematically underestimated in Pythia by a factor of $\sim2.8$ compared to the LEP measurement $f(b \rightarrow \overline{\Lambda}_b) = 0.101^{+0.039}_{-0.0031}$~\cite{Abbiendi_et_al__1999, PDG1998}). Our correction procedure works as follows: we calculate the prompt and $\bar{\Lambda}_b$-induced antinucleus fluxes separately with Pythia (in the default configuration). Before we add these contributions together (in order to obtain the total dark-matter-induced antinucleus formation) we rescale the $\bar{\Lambda}_b$-induced flux by a factor of 2.8. In this way we precisely compensate the underproduction of $\bar{\Lambda}_b$ particles in default Pythia.\footnote{Another correction procedure called the $\Lambda_b$-tune was proposed in~\cite{Winkler:2020ltd}. In the $\Lambda_b$-tune, the diquark formation rate in Pythia is adjusted to reproduce the correct transition ratio $f(b \rightarrow \overline{\Lambda}_b)$ measured by LEP. However, the change of the diquark formation rate also affects the production of antinucleons in Pythia. While this is not a problem as long as one correctly adjusts the coalescence momentum such that experimentally observed antinucleus formation rates are reproduced~\cite{Comment}, some misinterpretation has occurred in the literature~\cite{Kachelriess:2021vrh}. In this work, in order to avoid any possible confusion, we decided to employ the simpler correction procedure explained in the main text.}
(see Ref.~\cite{DelaTorreLuque:2023vvo} for the direct comparisons of the spectra from DM signals with and without this correction).



\paragraph*{Coalescence Momentum}
$\,$\\
The coalescence momentum entering our calculation of antinucleus fluxes is extracted from fits to the available experimental data. 
In fact, we would in principle need to distinguish four different coalescence momenta depending on (i) whether we are dealing with dark matter annihilation (in the event-by-event coalescence model) or with pp-collsions (in the analytic coalescence model), (ii) whether we are calculating antideuteron or antihelium formation.\footnote{The coalescence momenta for antideuteron and antihelium formation are expected to be different due to quantum mechanical arguments (see e.g.~\cite{Blum:2019suo}).} 
Luckily our fits below suggest very similar $p_c$-values for pp- and $e\bar{e}$-collisions (used as a proxy for dark-matter annihilation), such that we can drop the distinction (i). Hence, it is sufficient to distinguish the coalescence momenta for antideuteron and antihelium in the following.

We first extract the antinucleus coalescence momenta for pp-collisions in the analytic coalescence model. Specifically we perform joined fits to the CERN ISR~\cite{Alper:1973my,CHLM:1975xae,British-Scandinavian-MIT:1977tan} antideuteron data as well as the ALICE antideuteron and antihelium data~\cite{ALICEdata2018,ALICE2022}. The antinucleon spectra entering the calculation are taken from the parameterization of Ref.~\cite{Winkler:2017xor} for CERN ISR, while the measured antinucleon spectra are used for ALICE.\footnote{Because of ALICE's large center-of-mass energy ($\sqrt{s}=7,13\:\text{TeV}$) an additional (transverse-momentum-dependent) correction factor needs to be applied to Eq.~\eqref{eq:general}. The latter accounts for the hard correlations (which are not included in the analytic coalescence model). The determination of this correction factor is described in detail in~\cite{Winkler:2020ltd}. We follow exactly the same procedure here.} 

In addition to pp-collisions, we consider the antideuteron formation measured in $e\bar{e}$-scattering by ALEPH~\cite{ALEPH}. The resulting $p_c$ in the event-by-event coalescence model (employing the Pythia event generator) can be taken from~\cite{Winkler:2020ltd}.

\begin{figure}[!t]
\centering
\includegraphics[width=1.\textwidth]{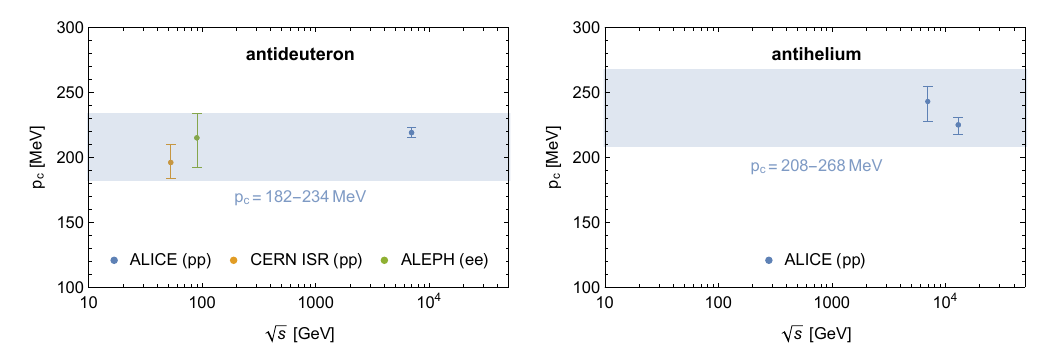} 
\caption{Antideuteron (left panel) and antihelium (right panel) coalescence momentum extracted from pp-scattering at CERN ISR~\cite{Alper:1973my,CHLM:1975xae,British-Scandinavian-MIT:1977tan} and ALICE~\cite{ALICEdata2018,ALICE2022}, and from $e\bar{e}$-scattering at ALEPH~\cite{ALEPH}. The blue uncertainty bands which we employ in this work amount to a conservative combination of the measurements.}
\label{fig:pc_fit}
\end{figure}

The antideuteron and antihelium $p_c$-values and uncertainties for CERN ISR, ALICE and ALEPH are depicted in Fig.~\ref{fig:pc_fit}. The shown uncertainty bands, which we employ in this work, amount to a conservative combination which encloses the $p_c$-measurements and $1\sigma$-uncertainties. Such a conservative approach is suggested by the fact that the only available data were either taken decades ago before the era of modern detectors, or -- in the case of ALICE -- at extremely high center-of-mass energies far beyond the energies relevant in cosmic-ray antinucleus formation. In the case of antihelium, since data in the energy-range relevant for cosmic rays are entirely missing, we added an additional $5\%$-uncertainty. The coalescence momenta employed in this work thus read,
\begin{equation}
p_c(\bar{d})=208\pm 26\:\text{MeV},\qquad
p_c(\overline{\text{He}})=238\pm 30\:\text{MeV},
\end{equation}
for both, astrophysical and dark-matter induced antinucleus formation.

\begin{figure}[!t]
\centering
\includegraphics[width=0.48\textwidth]{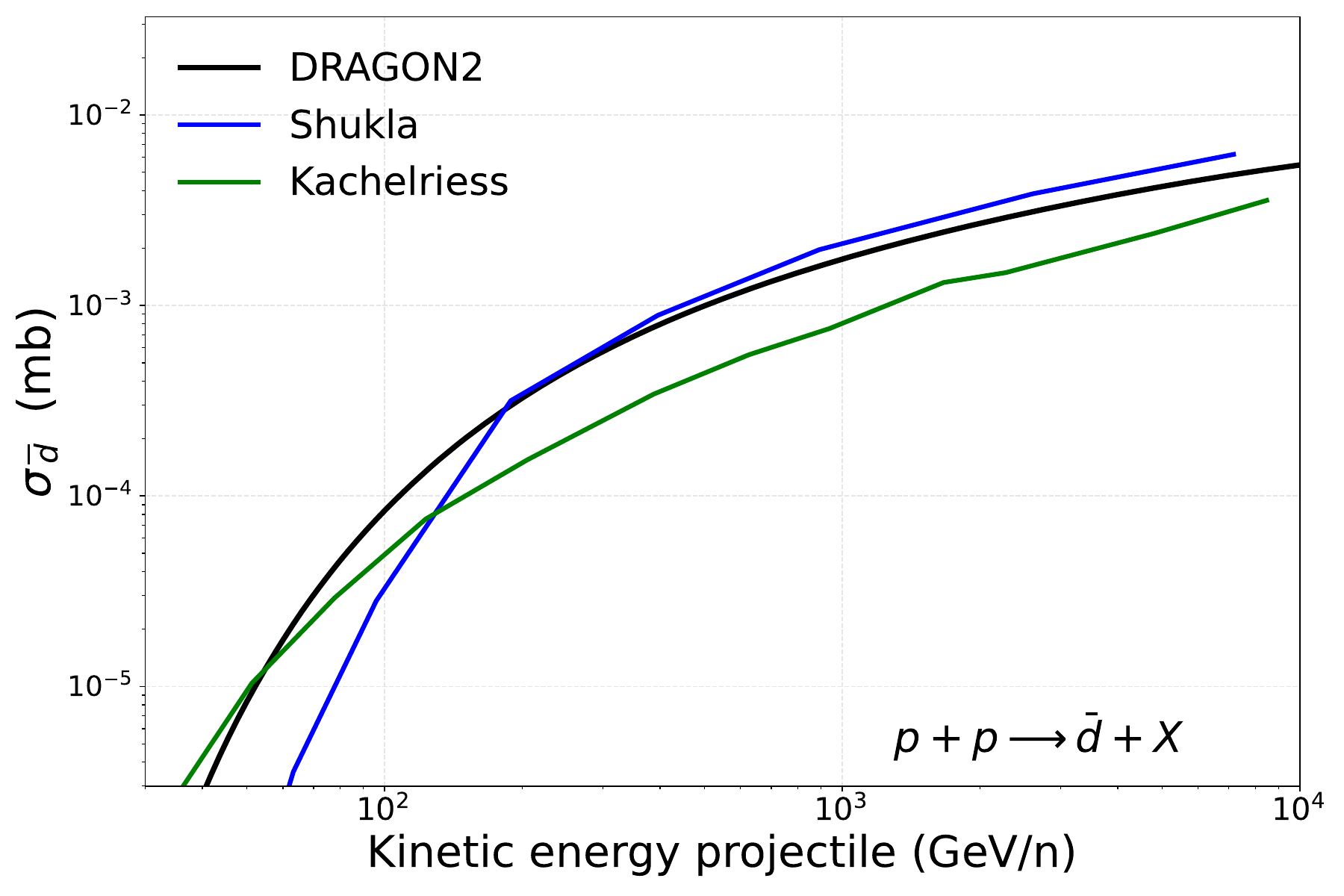} 
\hspace{0.1cm}
\includegraphics[width=0.48\textwidth]{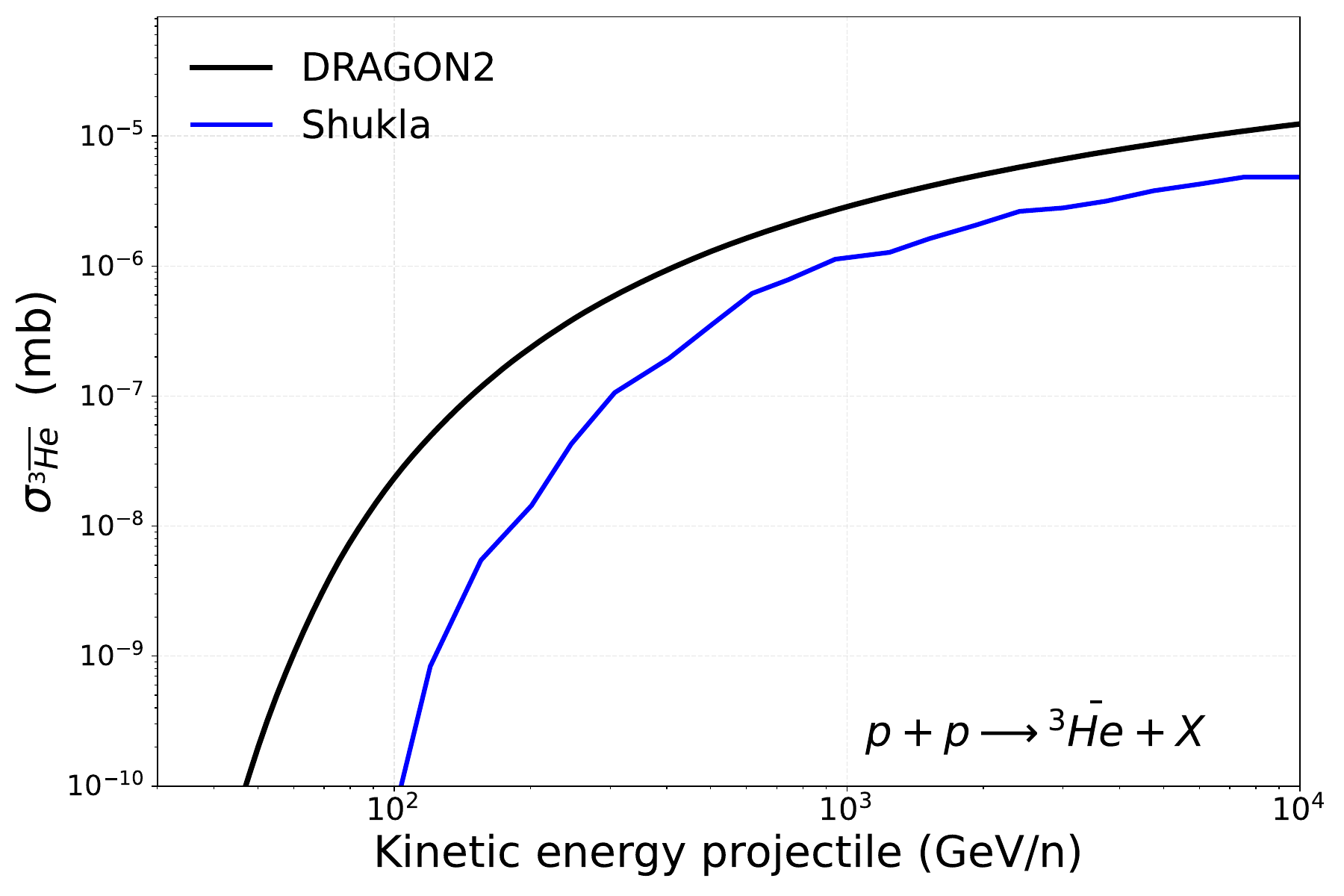}
\caption{Integrated total $\overline{d}$ (left panel) and $^3\overline{He}$ (right panel) production cross sections from $p+p$ collisions as a function of the projectile energy per nucleon. We compare our cross sections with those obtained by Shukla and collaborators~\cite{Shukla} and, for $\overline{d}$, also with those obtained by Kachelriess et al.~\cite{M_Kachelrie__2020} for $p+p$ interactions.}
\label{fig:SecAntinuclei}
\end{figure}

\paragraph*{Antinuclei Cross Sections and Production Spectra}
$\,$\\
In the left panel of Figure~\ref{fig:SecAntinuclei} we show the integrated total production cross sections for $\overline{d}$ production from $p+p$ collisions as a function of the projectile energy per nucleon, compared to those obtained by Shukla and collaborators~\cite{Shukla} and those obtained by Kachelriess et al.~\cite{M_Kachelrie__2020}. 

Our cross sections are similar to those of Shukla et al.\ above $200$~GeV, while they are around a factor of $2$ above those derived by Kachelriess et al. In this energy range, all calculations show a similar energy-dependence. However, at lower energies, these calculations differ more significantly. 
The right panel of this figure shows the integrated total production cross sections for $^3\overline{He}$ from $p+p$ collisions as a function of the projectile energy per nucleon, compared to those from Shukla et al. 

\begin{figure}[!t]
\centering
\includegraphics[width=0.48\textwidth]{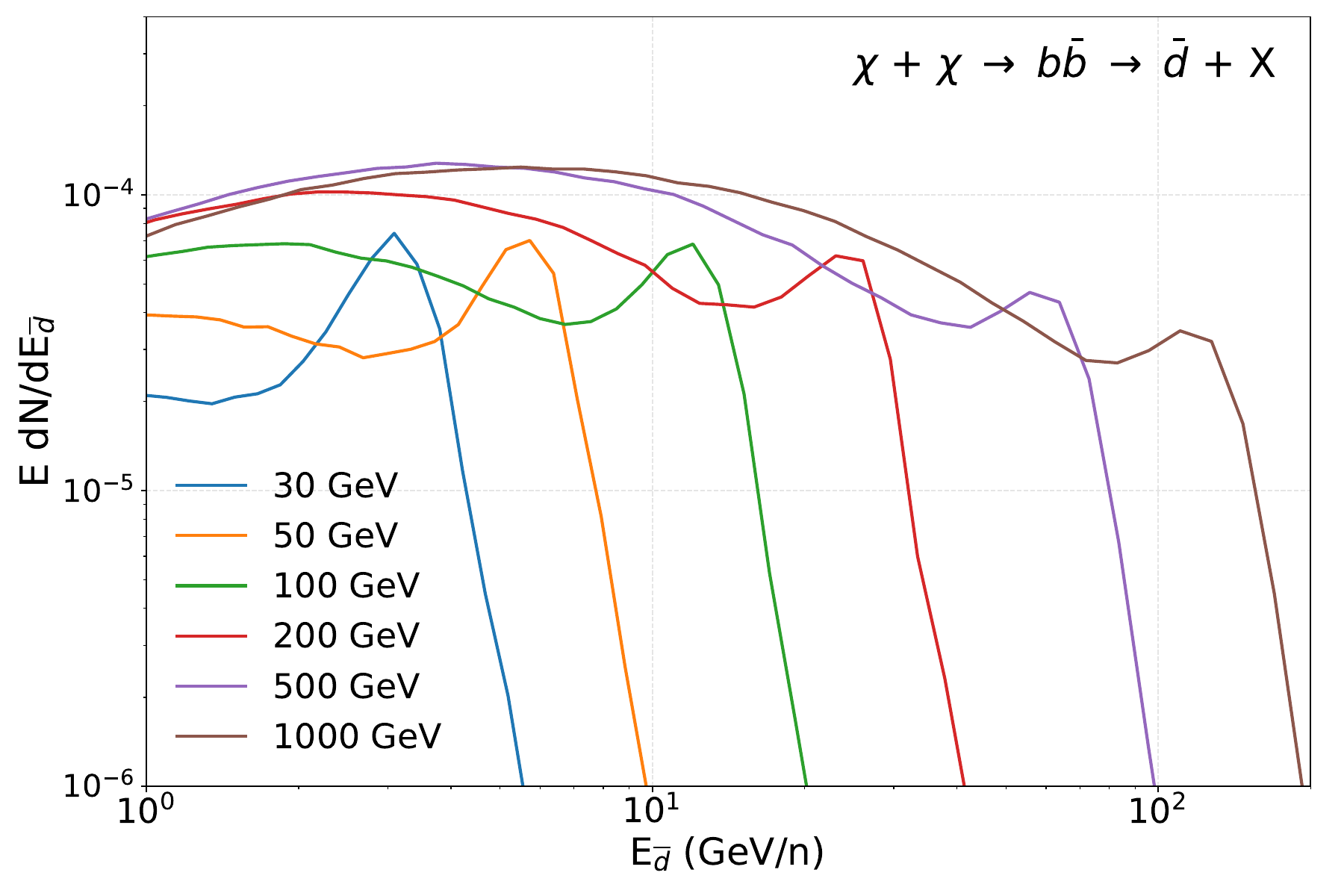} 
\hspace{0.1cm}
\includegraphics[width=0.48\textwidth]{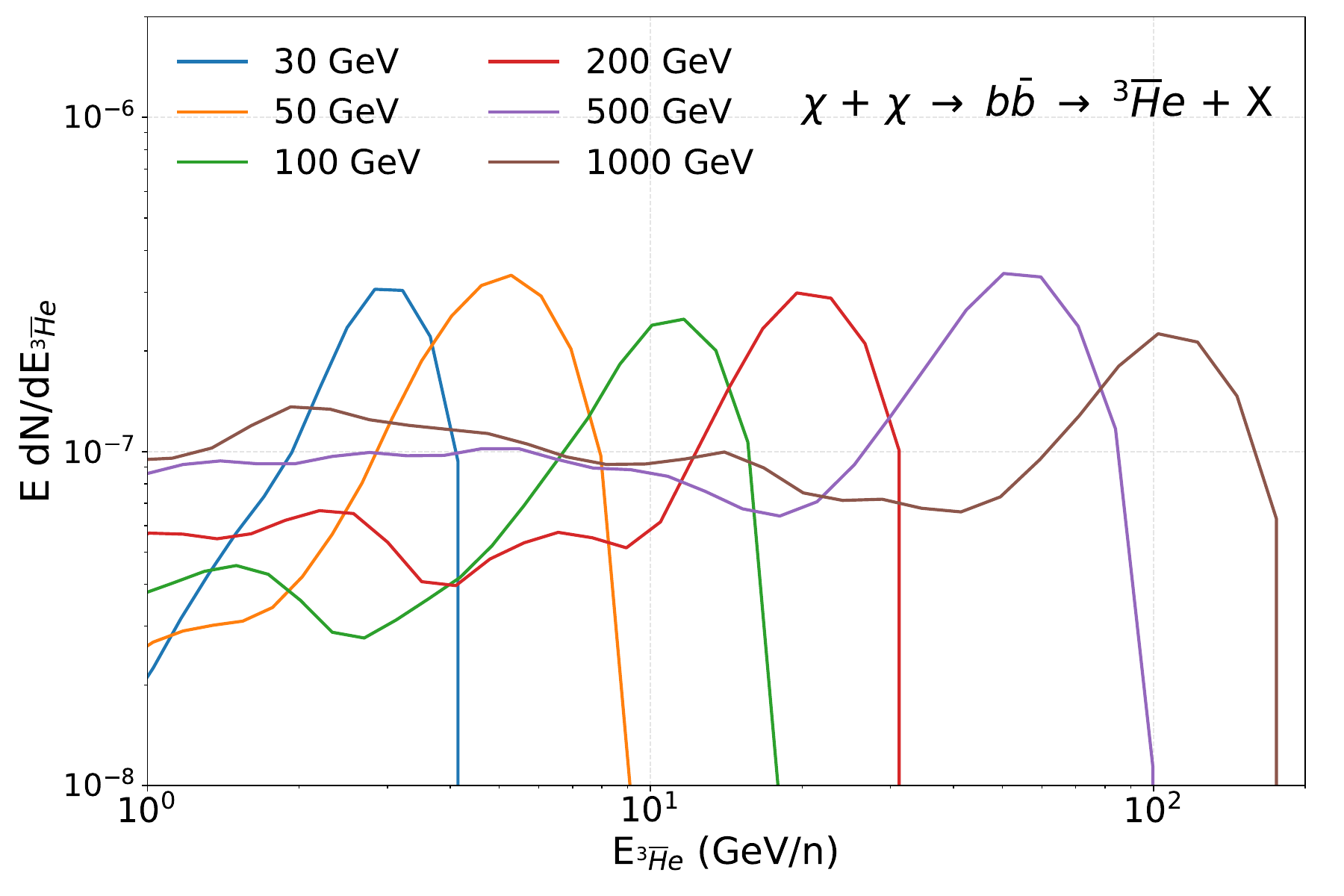}
\caption{Antinuclei spectra at injection produced by annihilation DM annihilation, in the $\bar{b}b$ channel, derived for WIMP masses of $30$, $50$, $100$, $200$, $500$ and $1000$~GeV for the production of $\overline{d}$ and $^3\overline{He}$.
\textbf{Left panel}: Computed $\overline{d}$ DM annihilation spectrum produced from a WIMP of different masses annihilating into $b\bar{b}$ final states, assuming $p_c=208$~MeV. \textbf{Right panel}: Same as in the left panel, but for $^3\overline{He}$, assuming $p_c=238$~MeV.}
\label{fig:DM_masses}
\end{figure}

In Figure~\ref{fig:DM_masses} we show the predicted $\overline{d}$ and $^3\overline{He}$ spectra (at production) arising from WIMP annihilation into $b\bar{b}$ final states for different WIMP masses. 
This figure clearly shows the bumps associated with the production of antinuclei from resonances (mainly $\overline{\Lambda}_b$ particles), which is particularly relevant for the production of $^3\overline{He}$ (right panel).

\section{Antinuclei implementation in {\tt DRAGON2}}
\label{sec:DRAGON}

{\tt DRAGON2}~\footnote{\label{note1}\url{https://github.com/cosmicrays/DRAGON2-Beta\_version}} is an advanced cosmic-ray propagation code, designed to self-consistently solve the diffusion-advection-loss equation that self-consistently describes cosmic-ray transport for all cosmic-ray species in the network. The code is capable of including both astrophysical and exotic sources (e.g., dark matter annihilations/decays). The transport equation features fully position- and energy-dependent transport coefficients in both two-dimensional (assuming cylindrical symmetry) and three-dimensional configurations of the Galaxy structure. {\tt DRAGON2} allows detailed study of both small-scale and large-scale structures (e.g., the spiral structure of the Galaxy) in steady-state and transient modes, refining the spatial resolution on the regions of interest (e.g., the local bubble, Galactic Center, or Galactic Plane).

Here, we describe the main aspects of the implementation of the $^3\overline{He}$ and $\overline{d}$ species in this version of the {\tt DRAGON2} code, which we have made publicly available at \url{https://github.com/tospines/Customised-DRAGON-versions/tree/main/Custom_DRAGON2_v2-Antinuclei}, together with the relevant cross sections tables derived in this work.

\subsection{Implementation of cross sections for secondary production}

To calculate the production of secondary nuclei and antinuclei, {\tt DRAGON2} utilizes a data-driven model for the cross-sections and energy spectra (i.e. d$\sigma$/dE, as a function of the projectile and daughter nuclei energy per nucleon) of the interactions $p+p$, $p+He$, $He+p$ and $He+He$, as well as $\bar{p}+p$ and $\bar{p}+He$. The relevant cross sections were derived with the analytic coalescence model as described in the previous section. To model the spallation and production of heavier (anti-)nuclei (e.g. C, N, O), {\tt DRAGON2} implements a simple scaling from the cross sections of $p+p$ interactions of the form $A^s$, where $A$ is the mass number of the progenitor nucleus and $s$ is a scaling index that can be set by the user for heavy nuclei in the CR network. The cross sections used for antinuclei production cross sections are those obtained as explained above. 

\subsection{Tertiary production and inelastic cross sections}
The fragmentation of heavier nuclei through their inelastic scattering off of the interstellar medium can be an important process at low energies, where our detectors will be more sensitive. In our implementation, the inelastic cross sections for antinuclei are obtained by scaling those of $\bar{p}+p$ described in Ref.~\cite{Tan_1983} and the $\bar{p}+$nucleus cross sections as parameterised in Ref.~\cite{MO97} (for more details, see section 7 of Ref~\cite{DRAGON2-2}). In this way, we estimate
the $\overline{d}$ inelastic cross sections as twice that of $\bar{p}$ at the same kinetic energy per nucleon, and the $^3\overline{He}$ cross sections as $3/2$ those of $\overline{d}$~\cite{Korsmeier_AN_2017}. A comparison with the most up-to-date inelastic cross sections measurements by Ref.~\cite{AntiHe3Ine, Ine_ALICE} showed that these parameterisations fall within data uncertainties above $0.1$~GeV.

An additional ingredient is the tertiary contribution, which can be dominant at energies below 1 GeV~\cite{Bergstrom_1999}.
Antinuclei also interact inelastically with the gas in the interstellar medium, undergoing annihilation, producing resonances like the $\Delta$ particle or losing a fraction of their energy (usually referred to as Non-annihilation inelastic cross sections)~\cite{DM_Antimatter}. These results in a contribution that can be dominant at low energies. This is accounted for in our implementation, following the procedure described in Ref.~\cite{DRAGON2-2}.

\subsection{Dark Matter Production}

The DRAGON2 code takes tables with the production spectrum (i.e. differential distribution, dN/dE) of each particle for the requested WIMP mass as an input and they are interpolated to the energies requested in the input file. This interpolation also smooths the distribution when the data given in the table is very disperse (for example, when there are uncertainties present in the spectrum calculation).

In addition to the dark matter annihilation spectrum, DRAGON2 includes configurable options to set the local dark matter density (at the solar position), the slope of the dark matter density profile, and the dark matter annihilation rate.

\section{Predicted antideuteron and antihelium spectra and prospects for future detection}
\label{Prospects}

In this section, we develop robust predictions for the expected fluxes of $\overline{d}$ and $^3\overline{He}$ at Earth based on the results obtained from the antiproton analyses carried out in our companion paper~\cite{Ap_analysis}. These analyses are based on a Markov Chain Monte Carlo procedure that determines the probability distribution functions of the CR propagation parameters entering in the diffusion coefficient (see Eq.2.1. of Ref.~\cite{Ap_analysis}), including reacceleration, the galactic halo height, WIMP mass and annihilation rate from a combined fit of the secondary CRs B, Be and Li together with antiprotons. Throughout this paper, we utilize our most detailed analysis, which we called the \textit{Canonical} model in Ref.~\cite{Ap_analysis}.

\subsection{Secondary production of antideuteron and antihelium}
\label{sec:astrop}

The secondary production of $\overline{d}$ and $^3\overline{He}$ is an important background, the impact of which we can evaluate thanks to the determination of the propagation parameters from other CR observables and the evaluation of the coalescence momentum, $p_c$, from recent accelerator data. Here, we show our evaluations using the best-fit propagation parameters obtained in our \textit{Canonical} analysis, in a scenario that includes no additional dark matter production. The propagation parameters inferred from the different antiproton analyses mentioned above are very similar, obtaining the same predictions. In fact, this is not a very relevant source of systematic uncertainty in these predictions, as was already shown by Ref.~\cite{Korsmeier_AN_2017}. 
Including inelastic cross section uncertainties were also found to have a very minor impact here~\cite{AntiHe3Ine}. In theory, the addition of non-uniform CR propagation could affect the interstellar spectra of the progenitors (mainly protons and helium) throughout the Milky Way.

\begin{figure}[!t]
\centering
\includegraphics[width=0.49\textwidth] {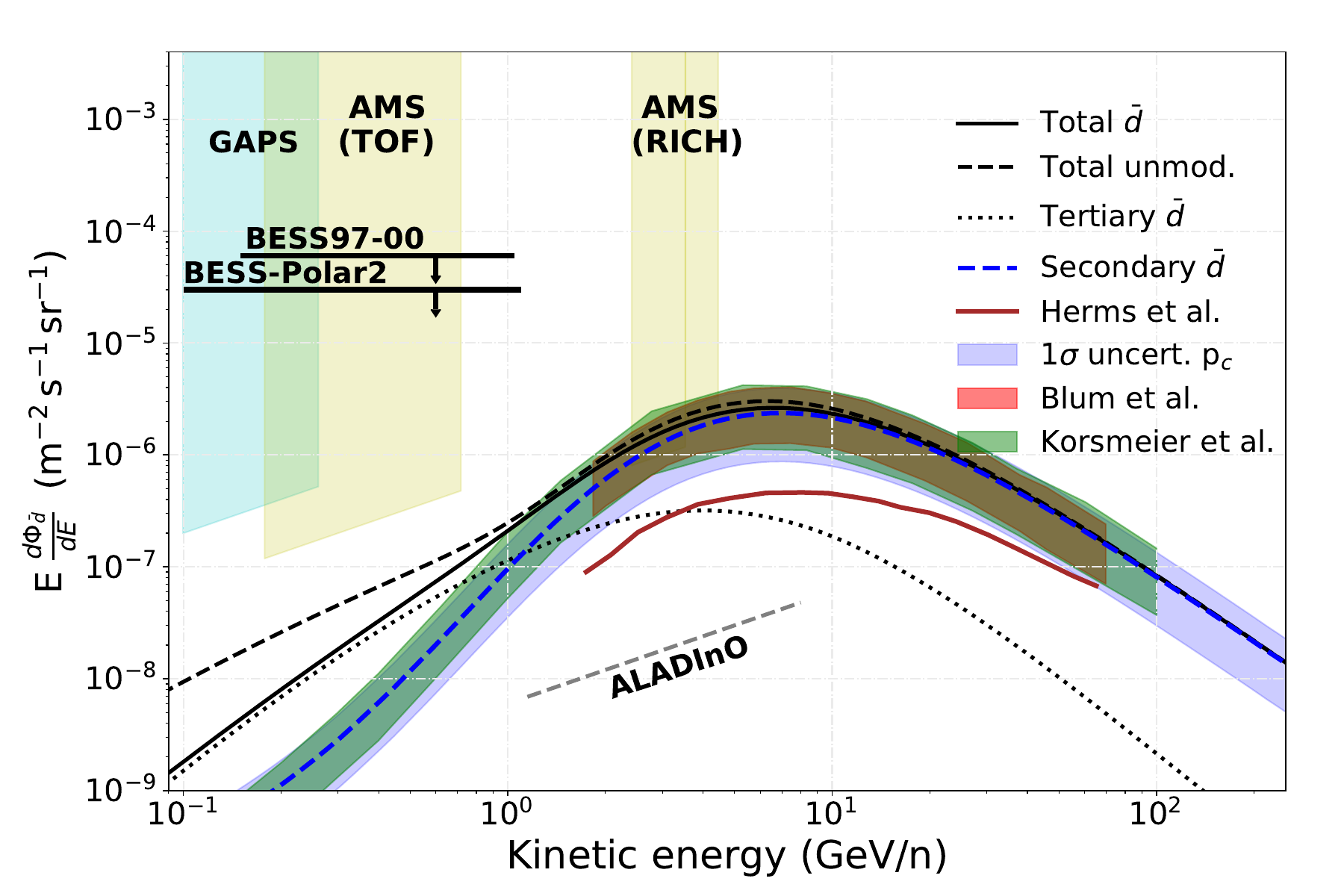} 
\hspace{0.1cm}
\includegraphics[width=0.48\textwidth] {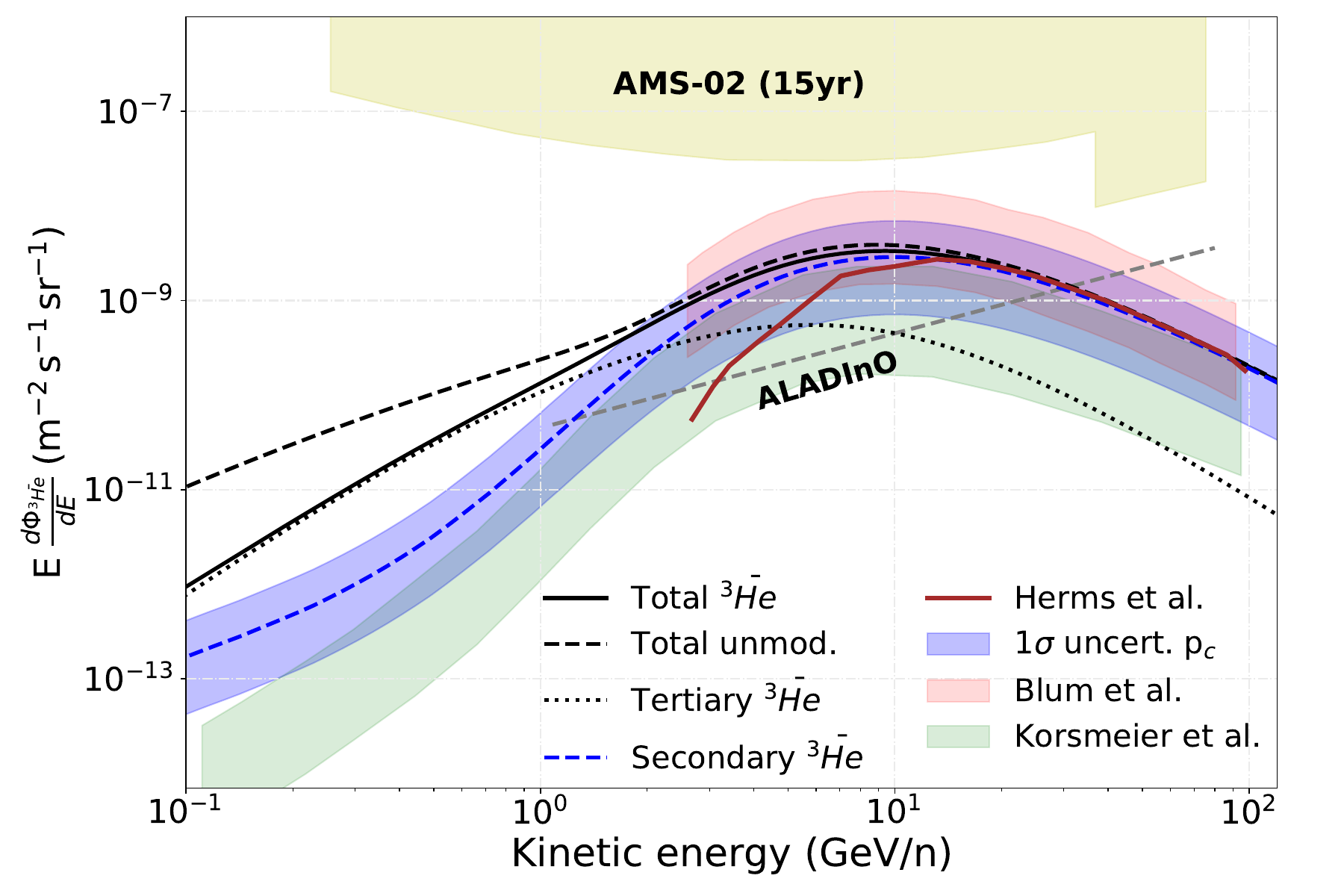}
\caption{\textbf{Left panel:} Predicted $\overline{d}$ spectrum produced from CR collisions on the interstellar gas (secondary $\overline{d}$) and the tertiary component, both modulated and unmodulated, compared to the upper-limits obtained by the BESS experiment, the sensitivity region of GAPS (three flights of $35$~days) and AMS-02 ($15$~years) and the ALADInO forecasted sensitivity in $5$~years of operation. \textbf{Right panel:} Similar to what is shown in the left panel, but for the $^3\overline{He}$ spectrum compared to AMS-02 ($15$~years) and the expected sensitivity of the future the ALADInO experiment. The uncertainty from the determination of $p_c=208\pm26$ (for $\overline{d}$) and $p_c=238\pm30$~MeV (see Fig.~\ref{fig:pc_fit}) is also shown in each panel. We employ the propagation parameters obtained in the {\it Canonical} antiproton analysis of Ref.~\cite{Ap_analysis}.
}
\label{fig:Astrophisical}
\end{figure}

Figure~\ref{fig:Astrophisical} shows the predicted $\overline{d}$ and $^3\overline{He}$ spectra produced by CR collisions with the interstellar gas and its tertiary component.
The dominant systematic uncertainty in our prediction is the coalescence momentum. We show how the uncertainty in antinuclei coalescence affects our predictions by including a band around the predicted secondary flux of these antinuclei (blue-dashed lines) with the central value found from the determination of the coalescence momentum in a fit to the existent data (see Fig.~\ref{fig:pc_fit}). In particular, this band corresponds to a value of $p_c = 208\pm26$ for $\overline{d}$ and $p_c=238\pm30$~MeV for $^3\overline{He}$.
In both panels, the total spectrum is modulated following the charge-dependent solar modulation approach~\cite{Cholis:2015gna} described in our companion paper. 

In the left panel of Figure~\ref{fig:Astrophisical}, we compare our prediction for the $\overline{d}$ spectrum to the upper-limits set from the Balloon-borne Experiment with a Superconducting Spectrometer (BESS)~\cite{BESS_upper}, the sensitivity regions for the RICH and TOF instruments in the AMS-02 detector (taken from Ref.~\cite{ARAMAKI20166}), for $15$~yrs of operation, and the expected sensitivity region for the General Antiparticle Spectrometer (GAPS)~\cite{ARAMAKI20166, GAPS_Ap} (assuming three flights of $35$~days). Moreover, we include the forecasted sensitivity for the future Antimatter Large Acceptance Detector In Orbit (ALADInO)~\cite{Battiston_2021} (for $5$~years of operation). 
In the right panel, we compare our predicted secondary $^3\bar{He}$ spectrum compared to the expected sensitivity band AMS-02 ($15$~years)~\cite{Winkler:2020ltd} and the expected sensitivity for the ALADInO experiment. Similar sensitivities to those forecasted for ALADInO are projected for the future AMS-100~\cite{AMS-100} experiment (expected to be operative after 2039).

Finally, in this figure, we compare our prediction for the secondary production with results from Ref.~\cite{Herms_2017} and the uncertainty bands obtained from Refs.~\cite{Korsmeier_AN_2017, Blum}~\footnote{In both cases, these bands represent uncertainties on the coalescence probability.}. As we see, our prediction is compatible with those from both Blum et al.~\cite{Blum} and from Korsmeier et al.~\cite{Korsmeier_AN_2017} (where the uncertainty band drawn corresponds to the uncertainty in the calculations with coalescence momentum from $160$~MeV to $248$~MeV and diffusion parameters from the {\it CuKrKo} model), although our model lies close to the upper-bound of the uncertainty bands. We note that the spectrum below a few GeV/n is significantly affected by the uncertainties in solar modulation, which are modelled in different ways for the different predictions compared here.
However, our prediction lays around one order of magnitude above that of Herms et al.~\cite{Herms_2017}, which from an event-by-event coalescence model using a modified version of the Monte Carlo generator DPMJET-III, opposite to the other calculations that relied on a semi-analytical approach. 

Excitingly, these results indicate that the detection of secondary antideuterons is achievable in a short term.
In fact, our predicted secondary antideuteron flux could explain the preliminary detection of $\mathcal{O}(1)$ antideuteron event by AMS-02. More precisely, we show that this contribution is expected to be detected by the RICH detector (intermediate-energy regime), while a detection by either the AMS-02 TOF system (in the low-energy regime) or by the GAPS experiment would provide evidence for new physics. Moreover, the ALADIno experiment should be able to perform detailed measurements of the $\overline{d}$ flux.

The predicted secondary production of $^3\overline{He}$, on the other hand, lies roughly one order of magnitude below the expected sensitivity of AMS-02 (for $15$~years of operation)~\cite{Cholis:2020twh}~\footnote{The AMS-02 collaboration does not report the detector sensitivity for $\overline{d}$ and $\overline{He}$, and thus we utilize sensitivity projections produced outside of the collaboration, as indicated in the main text.}, as shown in the right panel of Fig.~\ref{fig:Astrophisical}. This means that, if the preliminary measurement of antihelium events from AMS-02 is confirmed, this detection likely requires new physics which either boosts the expected secondary production or which constitutes a new source of antihelium. One example of new physics, a WIMP annihilating into bottom quarks, will be investigated in the next section.

In the right panel of Figure~\ref{fig:Astrophisical} we also display the expected sensitivity for the ALADInO experiment, finding that it would be capable of taking detailed measurements of the $^3\overline{He}$ spectrum in the energy region from $\sim 1$~GeV/n to $\sim30$~GeV/n.
We also compare our results with predictions from  Refs.~\cite{Korsmeier_AN_2017, Blum, Herms_2017}. As we see, our prediction is compatible with the uncertainty band from Blum et al.~\cite{Blum}, while  it lies slightly above the one from Korsmeier et al.~\cite{Korsmeier_AN_2017} above $\sim2$~GeV/n. We also notice that the predicted flux from Herms et al.~\cite{Herms_2017} follows an energy trend quite different from the one predicted by the other models below $\sim20$~GeV/n, but it becomes compatible with our predictions above that energy.

We also remark that our uncertainty band for antihelium is slightly smaller than that of previous works, which were released before the ALICE data became available (which are needed to constrain the antihelium coalescence momentum).

\subsection{Antinuclei from Dark matter Annihilation}
\label{sec:DM}

In this subsection, we present the expected fluxes of $\overline{d}$ and $^3\overline{He}$ produced by WIMP annihilation. We emphasize that we will only consider $\bar{b}b$ as a final state. In our predictions, the main uncertainties come from the coalescence model and from the branching ratio $\bar{\Lambda_b}\rightarrow\bar{d},\overline{He}$. While we will illustrate the coalescence uncertainty by drawing bands around our prediction which correspond to range of $p_c$ extracted in Sec.~\ref{sec:Coalescence}, it is very difficult to assess the uncertainty in the mentioned branching ratio.  This is because the processes $\bar{\Lambda}_b\rightarrow\bar{d},\overline{He}$ has not yet been measured experimentally and we can thus only extract its branching ratio from our Monte Carlo simulation (which could easily be off by an order of magnitude).

Finally, we note that the annihilation rate and mass of the dark matter particle are unknown. Therefore, we show predictions for either the most significant DM candidate found in our antiproton analysis or the maximal antinuclei fluxes allowed by our antiproton upper limits for different masses.

\begin{figure}[!t]
\centering
\includegraphics[width=0.495\textwidth, height=0.23\textheight]{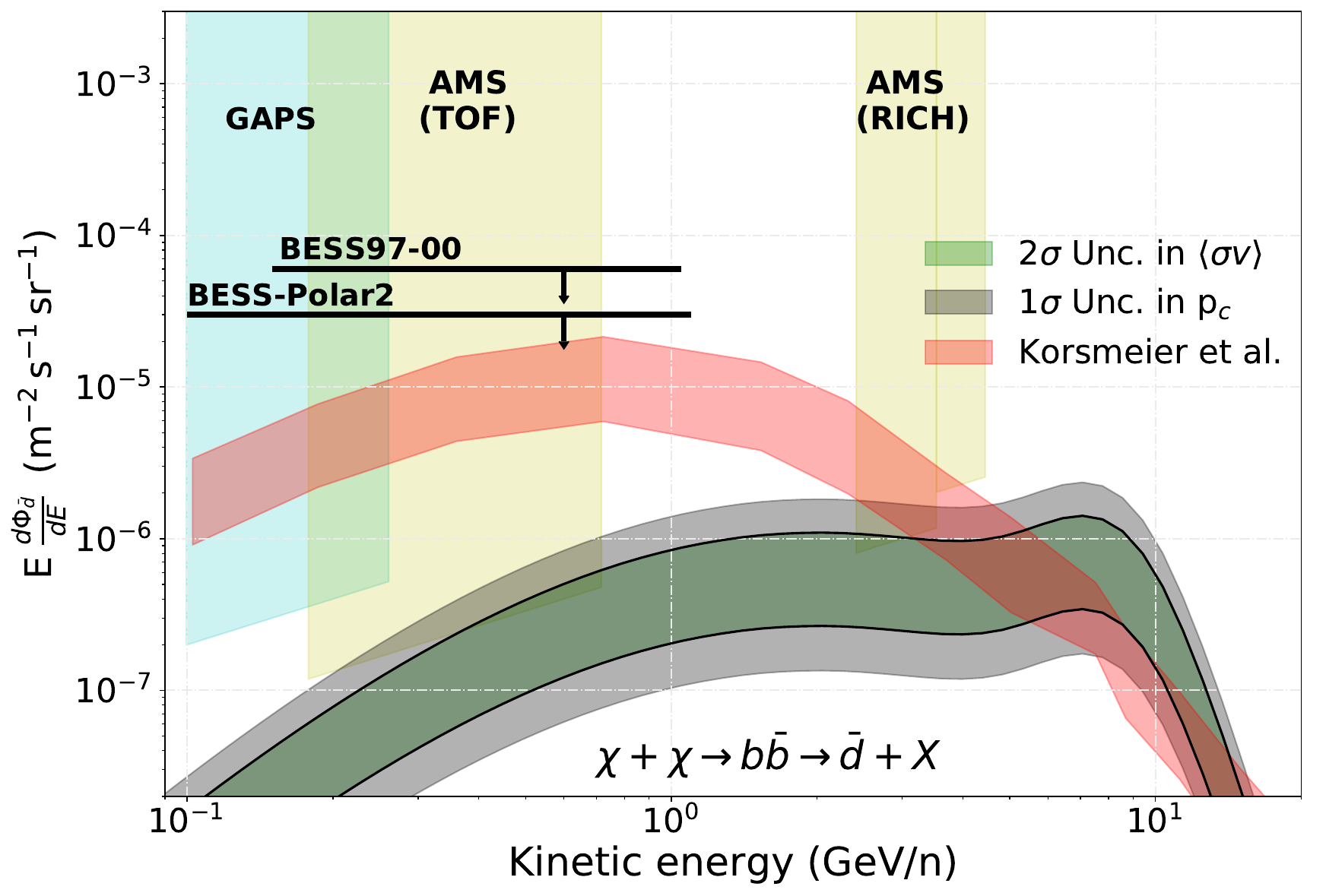} 
\hspace{0.1cm}
\includegraphics[width=0.475\textwidth, height=0.23\textheight]{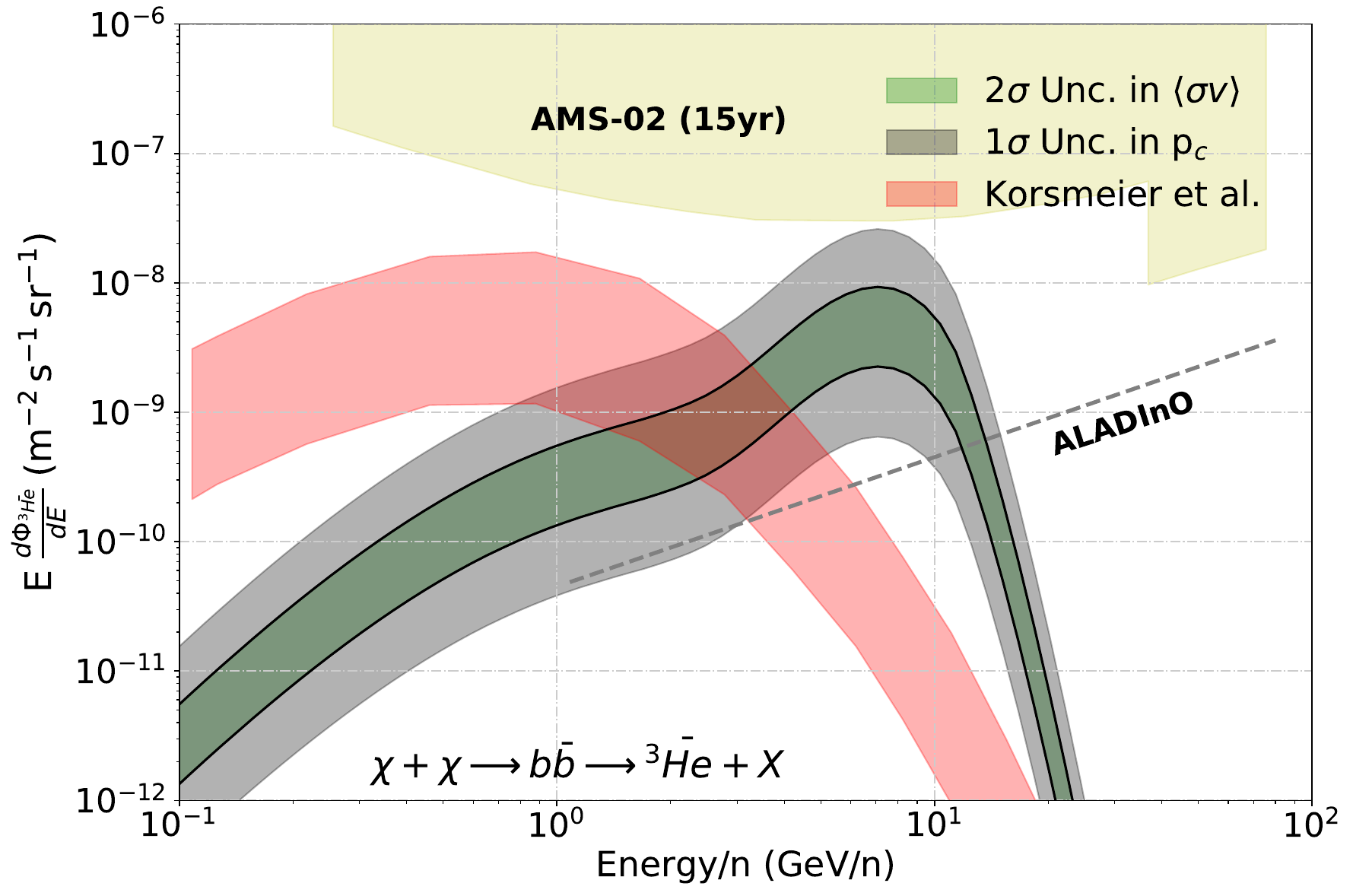}
\caption{Predicted antinuclei spectrum produced from annihilation into b$\bar{b}$ states of the best-fit WIMP found in our combined analysis~\cite{Ap_analysis} (m$_{\chi} = 66.28$~GeV, $\langle\sigma v\rangle= 0.99$~cm$^3$/s). The green uncertainty band corresponds to the $2\sigma$ uncertainty found in $\langle\sigma v\rangle$, while the uncertainty in the estimation of p$_c$ is shown as a grey band. These spectra are modulated as explained in the text. 
In the left panel we show the predicted $\overline{d}$ spectra compared to the upper-limits obtained by the BESS experiment, the sensitivity region of GAPS (three flights of $35$~days) and AMS-02 ($15$~years) and the ALADInO forecasted sensitivity in $5$~years. In the right panel we show the predicted $^3\overline{He}$ spectra compared to the expected sensitivity for AMS-02 ($15$~years) and the future the ALADInO experiment ($5$~years). 
In addition, we compare in both panels with the predicted DM signal from antiproton analyses used by Ref.~\cite{Korsmeier_AN_2017}.
}
\label{fig:DM_models}
\end{figure}

First, we focus on the WIMP parameters found with highest statistical significance in our Canonical antiproton analysis, which are m$_{\chi}=66.28$~GeV and $\langle \sigma v \rangle \sim  10^{-26}$~cm$^3$/s~\cite{Ap_analysis}. We emphasize, however, that the global significance associated with this WIMP candidate was below $2\sigma$ in our Canonical analysis, and we, hence, merely take these parameters as a benchmark.
The predicted local spectrum of $\overline{d}$ and $^3\overline{He}$ from such a WIMP annihilating in the b$\bar{b}$ channel are shown in Figure~\ref{fig:DM_models}, where we include a band related to the $2\sigma$ uncertainty in the determination of $\langle \sigma v \rangle$ in our analysis and another band representing the uncertainties related to the coalescence momentum. We also show the predictions from Ref.~\cite{Korsmeier_AN_2017}, where the authors used the analytic coalescence model for predicting DM-induced fluxes, rather than the event-by-event model. This, as explained above, neglects the correlations between the antinuleons produced (thus overestimating the low-energy antinucleus flux and underestimating the high-energy flux). Furthermore, the analytic model entirely misses the $\overline{\Lambda}_b$-induced contribution, which is the dominant component at high energies.
This emphasizes the importance of using a consistent coalescence treatment for these predictions and explains the important difference between our evaluation and those using an analytical coalescence treatment (totally erroneous for estimations of DM-induced antinuclei fluxes). In particular, this comparison allows us to see that DM signals that were allowed by some previous antiproton analyses predicted an antideuteron flux measurable by the current detectors, while our updated prediction shows that DM identification by antideuteron measurements are not easily achievable by the current experiments.
In these predictions, we simulate the effect of solar modulation with a modified Force-Field~\cite{forcefield}, which accounts for the effects of charge-sign dependence~\cite{Cholis:2015gna}, as in our antiproton analysis. 
We emphasize that uncertainties in the evaluation of the effect of solar modulation can also be very important for the spectra produced by WIMPs with mass $\lesssim 50$~GeV, since they peak at energies below a few GeV/n.

We remind the reader that in these computations we account for antinuclei produced at displaced vertices and, in addition, we correct for the branching ratio of production of $\overline{\Lambda}_b$ particles to reproduce the LEP measurements. As we observe in the left panel of Fig.~\ref{fig:DM_models}, the bump at around $10$~GeV/n, produced from the decay of $\overline{\Lambda}_b$ particles formed from the annihilation of the WIMP particle might be difficult to detect in the total $\overline{d}$ spectrum, given the high $\overline{d}$ background expected from CR interactions at those energies (left panel of Fig.~\ref{fig:Astrophisical}). As we see from the figure, the total DM signal could be detectable by the RICH and TOF detectors of AMS-02, being more difficult to be detected by GAPS, although not unfeasible given the large expected uncertainties.

However, in the case of $^3\overline{He}$, the bump produced by $\overline{\Lambda}_b$ particles would manifest very clearly, constituting a promising feature to reveal the existence of these signals from WIMP annihilation. 
But still, the expected flux lies slightly below the sensitivity of AMS-02, indicating some challenge in explaining the tentative detection of a few  $^3\overline{He}$ events by AMS-02 with a WIMP annihilating to $\bar{b}b$. 
However, it is difficult to draw any definite conclusion due to the lack of experimental data on the branching ratio $\bar{\Lambda}_b\rightarrow\overline{He}$ which introduces an additional major source of uncertainty beyond the one shown in Fig.~\ref{fig:DM_models}. Therefore, the measurements of the process $\bar{\Lambda}_b\rightarrow\overline{He}$ at LHC are eagerly awaited.
Nonetheless, it is important to state that this mechanism cannot explain the detection of $^4\overline{He}$, since it is not kinematically possible to produce $^4\overline{He}$ from this resonance. 

\subsubsection{Flux upper limits for antideuteron and antihelium from WIMP annihilation}
\label{sec:Maximal}

\begin{figure}[!t]
\centering
\includegraphics[width=0.49\textwidth]{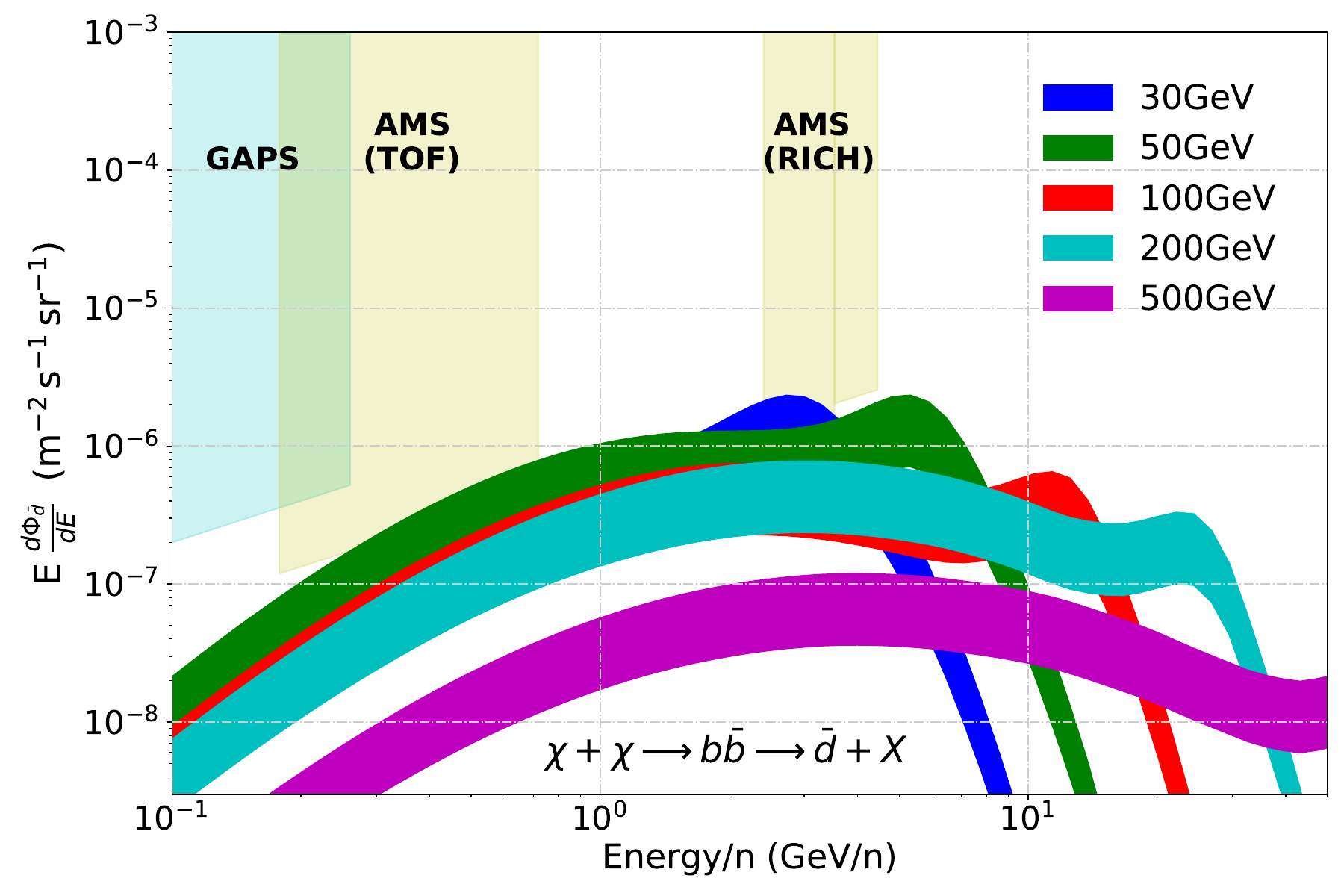} 
\includegraphics[width=0.49\textwidth]{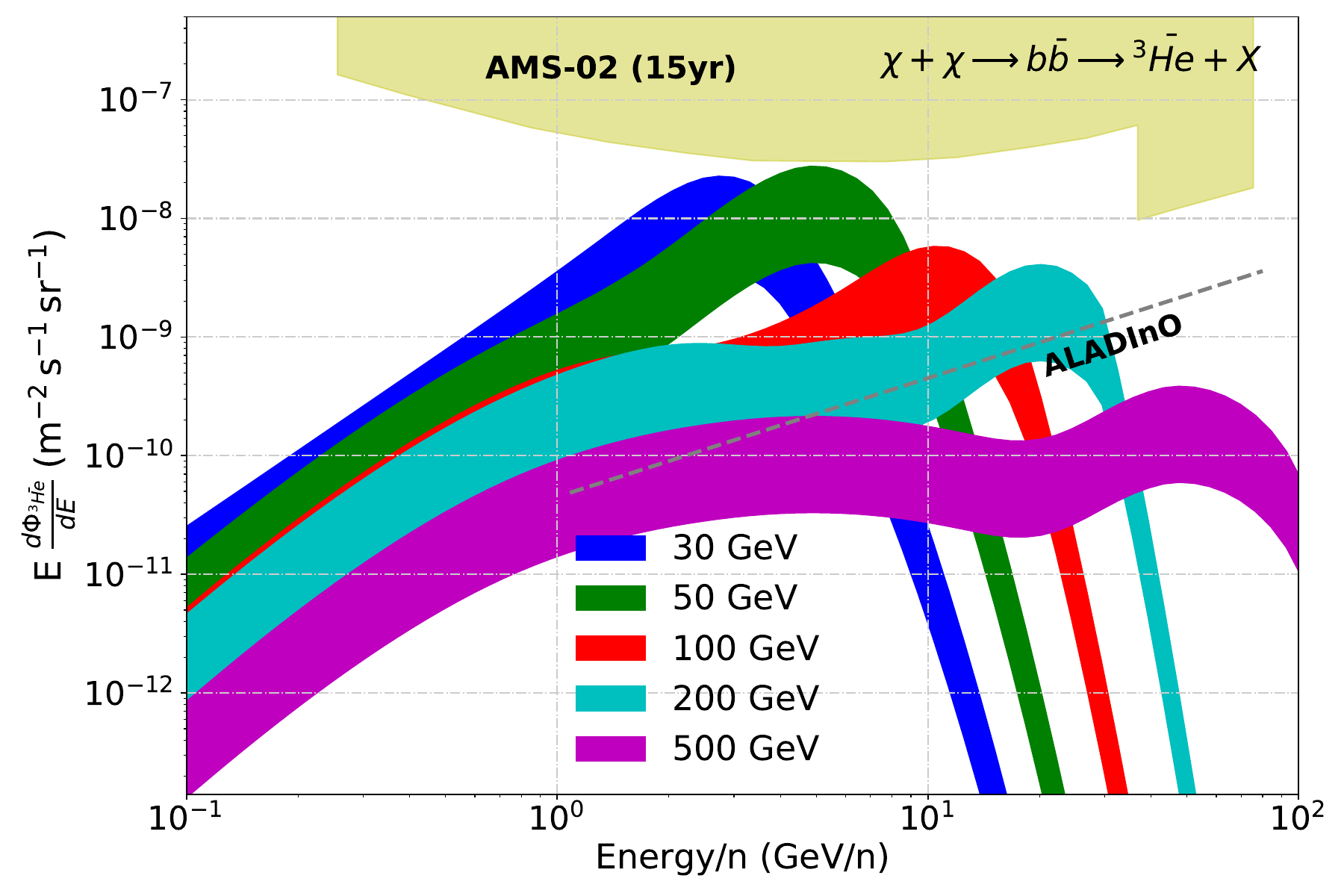} 
\caption{Upper limits on the antinuclei spectrum produced from the annihilation of a WIMP with different masses evaluated at the $95\%$ upper limit annihilation rate corresponding to the dark matter bounds obtained in our antiproton Analyses. The bands represent the $1\sigma$ uncertainty on p$_c$, as explained in the text.
\textbf{Left panel}: Predicted upper limits for the $\overline{d}$ spectrum from WIMP annihilation, compared to the sensitivity limits from GAPS and AMS-02.  \textbf{Right panel}: Predicted upper limits for the $^3\overline{He}$ spectrum from WIMP annihilation. Here, our predictions are compared to the expected sensitivity AMS-02 ($15$~years) and those for the future the ALADInO experiment ($5$~years).}
\label{fig:AntiHe_Max}
\end{figure}

Given the importance of a possible measurement of the antideuteron flux in the next years and the fact that our predictions seem to predict an antihelium flux which is slightly below the AMS-02 sensitivity, we report the maximum possible contribution to the production of $\overline{d}$ and $^3\overline{He}$ from WIMP annihilation that is allowed by our antiproton analyses. We will again only consider $\bar{b}b$ as a final state. In Figure~\ref{fig:AntiHe_Max}, we show the predicted maximal fluxes of $\overline{d}$  (left panel) and $^3\overline{He}$ (right panel) from WIMPs of different masses annihilating into $b\bar{b}$ final states, which keep the CR transport parameters and WIMP properties compatible with the current measurements of B, Be, Li and antiprotons.

In particular, Figure~\ref{fig:AntiHe_Max} shows the predicted spectra of these antinuclei produced from the annihilation of a WIMP with masses of $30$, $50$, $100$, $200$ and $500$~GeV evaluated at the $95\%$ upper limit annihilation rate in a NFW profile (corresponding to the dark matter bounds reported in the left panel Figure~3 of our companion work~\cite{Ap_analysis}) obtained in our Canonical analysis. How these signals change for a contracted-NFW profile is shown in Ref.~\cite{DelaTorreLuque:2023vvo}. 
As we see from the left panel, the expected $\overline{d}$ signal generated from WIMPs of masses below $\sim50$~GeV could be detectable by AMS-02 (both, in the TOF and RICH instruments) considering the uncertainties related, while GAPS is not expected to detect an event from these signals. 
We remind the reader, however, that the solar modulation parameters entering our flux prediction were adjusted to the AMS-02 data taking period.
Adjusting solar modulation to the period in which GAPS is expected to operate could slightly increase the predicted sub-GeV flux. However, even with this increase a detection at GAPS remains challenging.

We note that the best chance to identify DM with antideuterons would be the detection of a low-energy signal (possibly by the ToF detector), where the astrophysical flux (secondary $\overline{d}$ production) is negligible.  Another feature that could allow us to distinguish a DM signal from the astrophysical background is the peak produced by the $\overline{\Lambda}_b$ particles. However, this peak will be somewhat hidden by the secondary flux (which likely dominates at the relevant energies), and, therefore, difficult to identify.

Then, from the right panel of this figure, we observe that the predicted fluxes of $^3\overline{He}$ would lie slightly below the AMS-02 expected sensitivity. 
In order to explain the preliminary AMS-02 events with DM annihilation in the $\bar{b}b$-channel, a coalescence momentum at the upper end of the uncertainty (and possibly a statistical upward fluctuation of the flux) is required. Alternatively, a larger $\bar{\Lambda}_b\rightarrow\overline{He}$ branching ratio compared to our baseline Monte Carlo prediction (a possibility which will be experimentally scrutinzed in the next years at LHC) could lead to an observable antihelium flux. On the other hand, the AMS-02 collaboration recently showed that the antihelium events have likely been detected at $\text{Energy/n}\gtrsim 10\:\text{GeV}$~\cite{PZuccon} which makes it even a bit more difficult to explain the signal by dark matter annihilation. Still, it was shown in~\cite{Winkler:2020ltd, Ding_2023} that slight variations of the DM-annihilation channel (for instance DM annihilation into $\bar{b}\bar{b}bb$ through an intermediate light mediator) can further significantly boost the antihelium flux and move the peak of the expected spectrum to higher energies compared to $\bar{b}b$. In this light, dark matter annihilation to antihelium through an intermediate $\bar{\Lambda_b}$ remains a viable option to explain a few $^3\overline{\text{He}}$ events.

We remark here that the maximal DM-induced antinuclei fluxes compatible with antiprotons would still be above the expected flux produced from CR secondary collisions up to masses $\geq500$~GeV at energies above $1$~GeV.
Additionally, it is important to highlight that the maximally allowed antinuclei fluxes are (roughly) independent of the halo height used in the antiproton analysis. This is because a different halo height would scale the flux of antinuclei and antiprotons in the same way.
For instance, a larger diffusion halo would make the antihelium flux for a given dark matter annihilation cross section larger. But at the same time, this larger halo would tighten the antiproton limits on the annihilation cross section, thus reducing the allowed antihelium flux. Both effects would cancel (almost) exactly and leave maximally allowed antihelium flux unchanged.

Finally, we note that annihilation of WIMPs in simple 2- or 4-body final states cannot produce any appreciable flux of $^4\overline{He}$. If a $^4\overline{He}$ signal by AMS-02 is confirmed, more general final states in the dark matter annihilation need to be considered. Arguably, the most reasonable 
option to produce an observable cosmic ray $^4\overline{He}$ signal is to employ high-multiplicity quark final states. The model in~\cite{AN_Cascade} employs heavy ($\sim$unitarity limit) dark matter particles that annihilate/decay in a QCD-like dark sector, producing a shower of dark particles (a ``soft-bomb''~\cite{Knapen:2016hky}) that subsequently decay via portal interactions into quark-pairs producing a high multiplicity final state. The latter gives rise to strongly enhanced $^4\overline{He}$-production. 

\section{Discussion and conclusions}
\label{sec:conc}

In recent years a tentative dark matter signal in the AMS-02 antiproton data has caused significant excitement in the cosmic ray community. However, the latest dedicated analyses have found the underlying excess to be insignificant, once all relevant uncertainties are included. In fact, these analyses revealed that the precision of the AMS-02 antiproton data has far surpassed the precision by which the astrophysical antiproton flux can be predicted -- making any search for a dark matter signal in the antiproton channel very challenging. 

In this light, cosmic ray antideuterons and antihelium constitute important complementary channels which potentially allow for a cleaner identification of a dark matter signal. The importance of the light antinuclei channels is further highlighted by the reported tentative observation of a few antihelium events at AMS-02, which has already triggered a number of studies investigating possible exotic explanations. 
At the same time, in the light of upcoming data, it is crucial to improve the predictions for the standard astrophysical antinuclei fluxes as well as for simple dark matter implementations.


In this work, we have, therefore, reported updated predictions of the $\overline{d}$ and $^3\overline{He}$ spectra from secondary CR interactions and from a generic WIMP annihilating into $b\bar{b}$ final states. For the prediction of the WIMP-induced fluxes we implemented the constraints arising from a simultaneous fit of the AMS-02 antiproton and secondary antinuclei (B, Be and Li) spectra. 
These allow us to compute the most up-to-date predictions for the expected spectra of these antinuclei, using the most recent CR and accelerator data, and compare them with other previous predictions in the literature.

To evaluate these fluxes we employ a new implementation of the propagation of these particles in the {\tt DRAGON2} code. This version of the code is made publicly available at \url{https://github.com/tospines/Customised-DRAGON-versions/tree/main/Custom_DRAGON2_v2-Antinuclei}. For this calculation, we have derived new sets of antinucleus production cross sections in $p+p$, $p+He$, $He+p$ and $He+He$ interactions from the parameterization of antinucleon cross sections~\cite{Winkler:2017xor}. Furthermore, we calculated new WIMP annihilation spectra with the Pythia Monte Carlo generator taking into account the most recent accelerator data. Additional measurements and studies at accelerators are crucial to improve the current and, often, oversimplified models of coalescence, which still constitute one of the main sources of uncertainty in our predictions. In addition, a dedicated study on the effects in the predicted spectra from including spatial dependence in the diffusion coefficient is left for future work.

From our evaluations, we highlight that already the secondary antideuteron flux could be detectable by the AMS-02 RICH detector at GeV energies. In addition, we find that the expected antideuteron flux produced by WIMP annihilation could be detectable by AMS-02 (both, in the RICH and TOF detectors) if the WIMP mass is below $\sim200$~GeV. In particular, the secondary flux and dark-matter induced antideuterons are both capable of explaining the hint for $\mathcal{O}(1)$ event at AMS-02 RICH, while a signal in the TOF would clearly point towards dark matter. At GAPS, a detection of an antideuteron signal appears more challenging.

In contrast, we find that the expected secondary antihelium flux falls below the present AMS-02 sensitivity by about one order of magnitude. On the other hand, standard WIMPs annihilating into $b\bar{b}$ final states could account for the detection of $\mathcal{O}(1)$ $^3\overline{He}$ events under optimistic assumptions. The dark-matter-induced antihelium signal is entirely dominated by the process $\bar{b}\rightarrow \bar{\Lambda}_b\rightarrow {^{3}}\overline{He}$ as first pointed out in~\cite{Winkler:2020ltd}. The unique feature produced in the antihelium spectrum by the decay of the $\overline{\Lambda}_b$ particle could be fundamental for future dark matter searches.
We further note that different annihilation channels can lead to an additional enhancement of the antihelium signal (as, for example, through light mediators, like in the $\chi \chi \rightarrow \bar{b}\bar{b}bb$ model commented above, that could lead up to an order of magnitude of signal enhancement). However, even these variations struggle to explain any possible observation of even a single $^4\overline{He}$ event due to kinematical limitations. 

Taken in whole, these results demonstrate that (standard) astrophysical mechanisms are highly unlikely to be capable of explaining the preliminary AMS-02 signal of a few antihelium events. If this signal is confirmed, production mechanisms stemming from dark mater annihilation through a $\bar{\Lambda}_b$ mediator remain the most reasonable explanation. However, even this explanation requires somewhat unexpected large (but still plausible) rates for anti-helium formation from $\bar{\Lambda}_b$ decay, which will be testable in near-future experiments. If the efficient production of anti-helium from $\bar{\Lambda}_b$ decays is ruled out (or additionally, if the observation of antihelium-4 is confirmed by future AMS-02 observations), then more exotic production mechanisms will need to be explored.

\acknowledgments

We are very grateful to Daniele Gaggero and Michael Korsmeier for valuable discussions related to the manuscript. This project used computing resources from the Swedish National Infrastructure for Computing (SNIC) under project Nos. 2022/3-27, 2021/3-42 and 2021/6-326 partially funded by the Swedish Research Council through grant no. 2018-05973. P.D.L. was supported by the Swedish Research Council under contract 2019-05135 during the first stages of this work and is currently supported by the Juan de la Cierva JDC2022-048916-I grant, funded by MCIU/AEI/10.13039/501100011033 European Union "NextGenerationEU"/PRTR, as well as the grants PID2021-125331NB-I00 and CEX2020-001007-S, both funded by MCIN/AEI/10.13039/501100011033 and by ``ERDF A way of making Europe''. P.D.L. also acknowledges the MultiDark Network, ref. RED2022-134411-T.

\bibliographystyle{apsrev4-1}
\bibliography{biblio}

\newpage
\appendix



\end{document}